\documentclass[a4paper,11pt]{article}
\pdfoutput=1
\usepackage{jheppub}

\usepackage[english]{babel}
\usepackage{amsmath}
\usepackage{graphicx}
\usepackage{color}
\usepackage{subfig}

\usepackage{ae,aecompl}
\usepackage[utf8x]{inputenc}
\usepackage{hyperref}

\newcommand{\Mpl}{M_{\text{pl}}}

\newcommand{\ddchi}{\ddot{\chi}_{\mathbf{k}}}
\newcommand{\dchi}{\dot{\chi_{\mathbf{k}}}}
\newcommand{\chik}{\chi_{\mathbf{k}}}

\newcommand{\ddX}{\ddot{X}_{\mathbf{k}}}
\newcommand{\dX}{\dot{X}_{\mathbf{k}}}
\newcommand{\X}{X_{\mathbf{k}}}

\newcommand{\ud}{\text{d}}
\newcommand{\x}{4\mu m t_1}
\newcommand{\etta}{\frac{\partial\ln\mu}{\partial\ln g}}
\newcommand{\ettatext}{\partial\ln\mu/\partial\ln g}
\newcommand{\ettatwo}{\frac{\partial^2\ln\mu}{\partial(\ln g)^2}}
\newcommand{\ettatwotext}{\partial^2\ln\mu/\partial(\ln g)^2}
\newcommand{\fnl}{f_{NL}}

\begin{document}

\title{Modulated preheating and isocurvature perturbations}

\author{Kari Enqvist and}
\author{Stanislav Rusak}

\emailAdd{kari.enqvist@helsinki.fi}
\emailAdd{stanislav.rusak@helsinki.fi}


\preprint{HIP-2012-23/TH}

\affiliation{Department of Physics and Helsinki Institute of Physics, University of Helsinki,
P.O. Box 64, FIN-00014 University of Helsinki, Finland}

\abstract{
We consider a model of preheating where the coupling of the inflaton to the preheat field is modulated by an additional scalar field which is light during inflation. We establish that such a model produces the observed curvature perturbation analogously to the modulated reheating scenario. The contribution of modulated preheating to the power spectrum and to non-Gaussianity can however be significantly larger compared to modulated perturbative reheating. We also consider the implications of the current constraints on isocurvature perturbations in case where the modulating field is responsible for cold dark matter. We find that existing bounds on CDM isocurvature perturbations imply that modulated preheating is unlikely to give a dominant contribution to the curvature perturbation and that the same bounds suggest important constraints on non-Gaussianity and the amount of primordial gravitational waves.
}

\maketitle

\section{Introduction}
\label{sec:intro}

One of the key issues for inflationary models is the question of how the inflaton energy is converted to radiation of ordinary particles. In the conventional theory of reheating \cite{Abbott:1982hn,Dolgov:1982th,Albrecht:1982mp} the inflaton decay is perturbative with an effective decay width $\Gamma$. In a typical model $\Gamma$ is much smaller than the Hubble parameter $H$ at the end of inflation so that at first the inflaton loses more energy to the expansion of the universe than is transferred to its decay products. Once the Hubble parameter decreases to $H=\Gamma$, reheating becomes efficient and the inflaton energy is transferred to relativistic particles. In the so-called modulated reheating scenario \cite{Dvali:2003em,Dvali:2003ar,Zaldarriaga:2003my} the inflaton decays into radiation also in the perturbative regime but with a width that is dependent on another, spatially fluctuating scalar field. This modulating field is assumed to achieve perturbations during inflation, giving thus rise to a space-dependent decay rate. 

In models with several scalar fields the decay of the inflaton can also take place in a non-perturbative regime through parametric resonance \cite{Traschen:1990sw,Kofman:1994rk,Kofman:1997yn} (for recent reviews see \cite{Bassett:2005xm,Allahverdi:2010xz}). In this case, the inflaton energy is converted to particles in violent bursts, followed by a later thermalization of the decay products. 

If the coupling of the inflaton to the preheat field depends on yet another light field, we would have modulated preheating: inflaton would decay non-perturbatively but the parametric resonance would be space-dependent. It has been argued that the modulation of effective, field-dependent coupling constants follows naturally from string theory, and the implications for preheating were discussed in \cite{Kofman:2003nx} and \cite{Podolsky:2005bw}. Curvature perturbations from modulated preheating were considered in \cite{Kohri:2009ac} but without the effects of backreaction. Modulated perturbations in the context of preheating in hybrid inflation were studied in \cite{Bernardeau:2004zz}, and in the context of trapped inflation in \cite{Matsuda:2007tr}.

If several fields are dynamical during inflation, in addition to the adiabatic perturbation there will  also arise an isocurvature perturbation. These can feed the curvature perturbation \cite{GarciaBellido:1995qq,Wands:2000dp,Gordon:2000hv}, but even if the additional fields do not evolve during inflation, the isocurvature perturbations from those fields can be converted into the curvature perturbation once inflation is over. This is the case for the curvaton scenario \cite{Enqvist:2001zp,Lyth:2001nq,Moroi:2001ct}, where the curvaton field may come to dominate the energy density of the universe after inflation (see \cite{Mazumdar:2010sa} for a recent review). This also holds for the modulated reheating scenario \cite{Dvali:2003em,Dvali:2003ar,Zaldarriaga:2003my} where the decay width of the inflaton depends on an additional field. In addition to isocurvature perturbations, another signature of multifield scenarios could be non-negligible non-Gaussianity that might be observable already by the Planck satellite.

In the present paper we consider an extension of the idea of modulated reheating to the theory of parametric resonance. We study the case where the coupling of the inflaton, denoted as $\phi$, to the preheat field $\chi$ is modulated by its dependence on an additional scalar field, denoted as $\sigma$. One possibility is that the modulating field $\sigma$ could be dark matter, and we investigate the implications of isocurvature constraints on such a model. We also study the non-Gaussianity and the dependence of the non-linearity parameter $f_{NL}$ on the modulating field.

The outline of the paper is as follows. In Section~\ref{sec:background} we briefly review the modulated reheating scenario and the theory of parametric resonance in an expanding universe. In Section~\ref{sec:modpre} we consider the effects of the modulated coupling in preheating and compute the power spectrum and non-Gaussianity. In Section~\ref{sec:isocurvature} we consider the constraints on the model from observational bounds on CDM isocurvature perturbations if the modulating field is responsible for dark matter. We conclude in Section~\ref{sec:conclusions} with a discussion of the results.

\section{Background}
\label{sec:background}
Let us begin by setting the stage for modulated preheating by introducing the notation and reviewing the ingredients needed for discussing modulated decay and parametric resonance.
\subsection{Modulated reheating}
\label{subsec:modre}

In the modulated reheating scenario, the decay width is inhomogeneous due to its dependence on a field $\sigma$ which was light during inflation and therefore acquired an almost scale-invariant spectrum of perturbations. Between the end of inflation and the time when $H=\Gamma(\sigma)$ the inflaton oscillates in a harmonic potential and the evolution of the universe is matter-like, $H\propto a^{-3/2}$, where $a$ is the local scale factor. After the energy is transferred from the inflaton to other particles at $H=\Gamma$, the universe becomes locally radiation dominated, $H\propto a^{-2}$. Due to the inhomogeneity of the transition some parts of the universe will have expanded more than others resulting in density, or alternatively, curvature perturbations on the hypersurface of constant density.

As is well known, the curvature perturbation on uniform density hypersurface can be defined in a gauge-invariant manner as \cite{Bardeen:1983qw,Bardeen:1989}
\begin{equation}
 \zeta = -\psi + \frac{1}{3(1+w)}\frac{\delta\rho}{\rho}.
\end{equation}
Similarly, for any non-interacting component $i$ we can define the curvature perturbation on the hypersurface where the energy density of that component vanishes as
\begin{equation}\label{zetai}
 \zeta_i = -\psi + \frac{1}{3(1+w_i)}\frac{\delta\rho_i}{\rho_i}~,
\end{equation}
so that the total curvature perturbation is given by
\begin{equation}\label{zetatot}
 \zeta = \sum_{i}\frac{(1+w_i)\rho_i}{(1+w)\rho}\zeta_i.
\end{equation}
For practical purposes, in modulated reheating we may treat the different regions as independent FRW-universes and make use of the $\delta N$ formalism \cite{Wands:2000dp,Lyth:2004gb}, where the curvature perturbation $\zeta$ is given by the difference of the expansion rate of separate universes with $\zeta=\delta N(\sigma)$. The local number of e-foldings after the end of inflation is given by
\begin{equation}
 N = \ln\left(\frac{a}{a_{end}}\right) = -\frac{2}{3}\ln\left(\frac{\Gamma}{H_{end}}\right) - \frac{1}{2} \ln\left(\frac{H}{\Gamma}\right).
\end{equation}
For modulated reheating the curvature perturbation on uniform energy hypersurfaces is then
\begin{equation}
 \zeta_{\text{MR}} = \delta N = N_{\sigma}\delta\sigma_* = -\frac{1}{6}\frac{\Gamma'}{\Gamma}\delta\sigma_* = -\frac{g'}{3g}\delta\sigma_*,
\end{equation}
where $\delta\sigma_*$ refers to field perturbation at the time of horizon crossing, and the last equality comes from assuming $\Gamma \propto g^2$ with $g=g(\sigma)$ a coupling constant. Here and in what follows $N_\phi$ is shorthand for $\partial N/\partial \phi$ for any field $\phi$.

\subsection{Parametric resonance in expanding universe}
\label{subsec:parres}

The simplest example of an effective potential giving rise to a parametric resonance is
\begin{equation}
\label{eq:potential}
V = \frac{1}{2}m^2\phi^2 + \frac{1}{2}g^2\phi^2\chi^2~,
\end{equation}
where $\phi$ is the inflaton and $\chi$ is the preheat field. The theory of parametric resonance in an expanding universe for such a quadratic model has been studied in detail in \cite{Kofman:1997yn}.
After the end of inflation, the inflaton oscillates in a harmonic potential around the minimum, and the evolution of the universe is matter-like. Every time the inflaton goes through the minimum of the potential the effective mass of $\chi$ becomes zero, and $\chi$ particles are copiously produced in a resonant regime, which is characterized by its resonance parameter
\begin{equation}
 \label{q0defined}
 q_0=\frac{g^2\Phi_0^2}{4m^2}~,
\end{equation}
where $\Phi_0$ is the amplitude of the inflaton at the end of inflation.

The number density of $\chi$ particles during the resonance grows as (see the Appendix):
\begin{equation}
 \label{eq:n_chi}
 n_{\chi}(t) \simeq \frac{(g\Phi_0 m)^{3/2}}{16 \pi^3 a^3\sqrt{2\mu mt}}e^{2\mu mt},
\end{equation}
where  $\mu$ is the effective Floquet exponent describing the growth rate of the fluctuations. The growing number of $\chi$ particles changes the masses of the fields, causing a backreaction which eventually shuts down the resonance. Backreaction becomes important when $n_\chi(t) \simeq m^2\Phi(t)/g$ \cite{Kofman:1997yn}. Comparing with eq.~\eqref{eq:n_chi} then gives the time at which this happens as
\begin{equation}
 \label{eq:end_time}
 mt_1 = \frac{1}{4 \mu}\ln\left[A^2\frac{m}{\Phi_0}\left(\frac{3H_0}{2m}\right)^2 \frac{(mt_1)^3 \mu}{g^5}\right],
\end{equation}
where we\footnote{Using the equations of~\cite{Kofman:1997yn} gives $A = 64 \pi^3$. However, it seems to us that \cite{Kofman:1997yn} has an error in their steepest descent formula. We have also used a different estimate for the location of the maximum in the Floquet index. See Appendix~\ref{sec:Floquet} for details.} find $A = 16\sqrt{2} \pi^{3}$. $H_0$ is the Hubble parameter at the beginning of the oscillation period so that $a(t) = (1 + 3H_0t/2)^{2/3} \simeq (3H_0t/2)^{2/3}$ at late times. This first stage of preheating is followed by a second stage where the effects of backreaction must be taken into account; however, as discovered by Kofman et al. \cite{Kofman:1997yn}, the second stage lasts a short while compared to the expansion rate of the universe. Hence we can consider the time $t_1$ to mark the end of preheating.

The produced particles are not in equilibrium. Lattice simulations show however that the effective equation of state $w = \left<p\right>/\left<\rho\right>$, where $p$ and $\rho$ are the pressure and energy density respectively, rises rapidly after a matter-like phase and levels off around a value that is typically somewhat below the radiation-like value $w=1/3$ \cite{Podolsky:2005bw, Frolov:2008hy,Sainio:2009hm,Easther:2010qz}.

\section{Modulated preheating}
\label{sec:modpre}
\subsection{Curvature perturbation}
\label{subsec:powerspec}
If the coupling $g$ of the inflaton to the preheat field $\chi$ in eq.~\eqref{eq:potential} depends on an additional scalar field $\sigma$ which is light during inflation, $g=g(\sigma)$, we expect that entropy perturbations in this field will cause a spatial variation in the process of preheating in analogy to the modulated reheating scenario. In the modulated reheating scenario, curvature perturbations are generated because of inhomogeneities in the time of transition of the equation of state from matter-like ($w=0$) to radiation-like ($w=1/3$). More generally, the scale factor grows as $a\propto t^{2/3(1+w)}$ so that if we have an instantaneous transition at some time $t_1$ from $w=w_i$ to $w=w_f$, the number of e-foldings is

\begin{equation}
 \label{eq:N_general}
 N = \ln{\left(\frac{a_1}{a_i}\right)} + \ln{\left(\frac{a}{a_1}\right)} = \frac{2}{3}\left(\frac{1}{1+w_i} - \frac{1}{1+w_f}\right)\ln{\frac{t_1}{t_0}} + \: \frac{2}{3}\frac{1}{1+w_f}\ln{\frac{t}{t_0}}
\end{equation}
If the transition is inhomogeneous, that is, if $t_1 = t_1(\sigma)$, curvature perturbations will be generated due to different expansion rates in different parts of the universe:

\begin{equation}
 \zeta_{\text{MP}} = \delta N = \frac{2}{3}\frac{w_f}{1+w_f}\frac{\partial \ln t_1}{\partial \ln g}\frac{g'}{g}\delta\sigma_*,
\end{equation}
where $w_i$ is assumed to vanish since the evolution during the period of oscillation of the inflaton is matter-like. Here $'$ denotes the derivative with respect to the field $\sigma$, and the subscript "MP" signals that the curvature perturbation contribution is due to modulation of preheating. In fact, as found by Podolsky et al. \cite{Podolsky:2005bw}, the final value of $w$ as well as the shape of the transition also depend on $g$ and therefore the last term in eq.~\eqref{eq:N_general} would also contribute to the curvature perturbation. Detailed investigation of this term would require a comprehensive study of thermalization that can only be done through extensive lattice simulations. Thermalization after preheating in a model with a quartic potential has been studied in \cite{Micha:2002ey,Micha:2004bv}. Here we assume that this effect is subdominant and that the subsequent evolution is independent of the coupling.

If the process of preheating terminates before backreaction becomes important, $t_1 \propto g$ \cite{Kofman:1997yn}, and the curvature perturbation is of the same order (but opposite sign) as that produced in the modulated reheating case. This case has been studied in \cite{Kohri:2009ac} including the loop contributions to the spectrum and to non-Gaussianity. We focus on the case where backreaction shuts down the resonance and restrict the analysis to tree level. We have checked that the tree contributions dominate for $\mathcal P_{\delta\sigma/\sigma}^{1/2} \lesssim 10^{-5}$ with the parameter values we have explored\footnote{For larger values of $\mathcal P_{\delta\sigma/\sigma}$ loop contributions may dominate with some tuning, however, these parameter values are ruled out by isocurvature constraints discussed in the next section.}. If backreaction shuts down the resonance then the end of preheating is given by eq.~\eqref{eq:end_time} and we get:

\begin{equation}
 \frac{\partial \ln t_1}{\partial \ln g} = -\frac{\x - 1}{\x-3}\left[\frac{5}{\x-1} + \etta\right].
\end{equation}
The duration of preheating, $mt_1$, is typically of the order of $10^2$ and typically $\mu \sim 0.13$. If the term $\ettatext$ is negligible, then the amplitude of the curvature perturbation is $\left|\zeta_{\text{MP}}\right| < 10^{-1}\left|\zeta_{\text{MR}}\right|$, where $\zeta_{\text{MR}}$ is the curvature perturbation in modulated reheating one would obtain with the same set of parameters. As a consequence, the contribution of modulated preheating to the power spectrum in this case is less than $10^{-2}$ relative to the corresponding modulated reheating scenario. 

\begin{figure}[!ht]
\begin{center}
\includegraphics[width=0.50\textwidth]{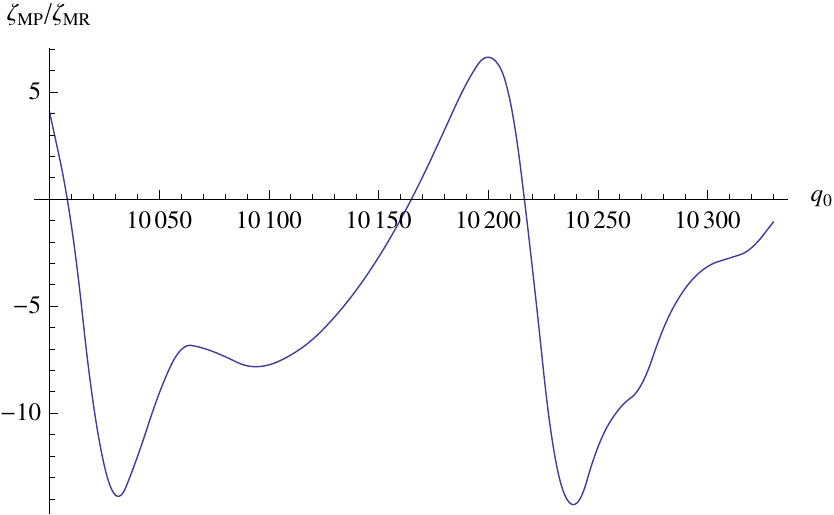}
\caption{Comparison of the curvature perturbation from modulated preheating with curvature perturbation from modulated reheating as a function of the resonance parameter $q_0$,  for $w_{f}=0.25, H_0 = 0.25m$. We have estimated $\ettatext$ by first solving the time at which the number of produced particles exceeds the limit where backreaction becomes important for each $q_0$ (to the accuracy of half an inflaton oscillation) and then solving the derivatives of $\mu$ at that time from eq.~\eqref{eq:n_chi}.}
\label{fig:ZetaMPMR}
\end{center}
\end{figure}

However, we find by solving numerically the equation of motion for the preheat field that the typical value of $\ettatext$  is $\mathcal O(10)$. Thus this term can be dominant. In this case $\left|\zeta_{\text{MP}}\right| \gtrsim 5\left|\zeta_{\text{MR}}\right|$ and the contribution of modulated preheating to the power spectrum can be significantly larger than in the modulated reheating scenario with a corresponding coupling profile. In Figure~\ref{fig:ZetaMPMR} we compare the curvature perturbation produced by modulated preheating to the curvature perturbation produced by the corresponding modulated reheating model in the broad resonance regime, $q_0 \gg 1$, where the interesting effects can be found. We take $q_0 \sim 10^4$, and as can be seen in Figure~\ref{fig:ZetaMPMR}, for a range in the resonance parameter $q_0$ of this order modulated preheating leads to a relatively high contribution.

\subsection{Non-Gaussianity}
\label{subsec:nongaussian}
The primordial curvature perturbation is given by the combined contributions from inflation and modulated preheating,

\begin{equation}
 \zeta = N_{\phi}\delta\phi_* + N_{\sigma}\delta\sigma_*,
\end{equation}
and the power spectrum, defined by $\mathcal{P}_{\zeta}(k)\delta(\mathbf{k-k'}) \equiv \frac{k^3}{2\pi^2}\left<\zeta_{\mathbf k}^*\zeta_{\mathbf k'}\right>$, is given by

\begin{equation}
 \mathcal{P}_{\zeta}(k) = \Big(N_{\phi}^2 + N_{\sigma}^2\Big)\left(\frac{H}{2\pi}\right)^2.
\end{equation}
Defining parameters

\begin{equation}
 \label{eq:param_definition}
 \lambda \equiv \frac{1}{2}N_{\sigma}\sigma_* = \frac{w_f}{1+w_f}\frac{g'\sigma_*}{3g}\frac{\partial \ln t_1}{\partial \ln g}, \qquad
 \xi \equiv 8\epsilon_*\left(\frac{\Mpl}{\sigma_*}\right)^2,
\end{equation}
we can write $\mathcal P_{\zeta} = (1+\xi\lambda^2)\mathcal P_{\text{inf}}$ with $\mathcal P_{\text{inf}} \equiv N_{\phi}^2(H/2\pi)^2$ being the contribution of the inflaton to the power spectrum. Here we have adopted the generic prediction $N_{\phi}^2 = 1/2\epsilon_*\Mpl^2$ where $\epsilon_* $ is the slow-roll parameter describing the deviation from deSitter space evaluated at the time of horizon crossing. We assume that the dependence of the coupling on the field $\sigma$ is of the form
\begin{equation}\label{formofg}
g^2(\sigma) = g_0^2(1+\sigma^2/M^2)~,
\end{equation}
where $M$ is some energy scale. According to our numerical estimates for $\ettatext$, the parameter $\lambda$ is at most $\mathcal O(1)$ and so modulated preheating can dominate the curvature perturbations only if $\xi$ is large, that is, if the amplitude of the field $\sigma$ during inflation is sufficiently below Planck scale.

Non-Gaussianity is given by the expression

\begin{equation}
 \frac{6}{5}\fnl = \frac{N^{a}N^{b}N_{ab}}{[N_cN^c  ]^2}~,
\end{equation}
where we have ignored the loop corrections. Non-Gaussianity in the case without backreaction including loop corrections was studied in \cite{Kohri:2009ac}.
The part coming from the inflaton is suppressed by slow-roll parameters and the only potentially large contribution is $N^{\sigma}N^{\sigma}N_{\sigma\sigma}/(N_{\sigma}N^{\sigma} + N_{\phi}N^{\phi})^2$. In our parameterization $N_{\sigma} ^2/N_{\phi} ^2 =\xi\lambda^2$ and

\begin{equation}
 N_{\sigma} = g'\frac{\partial N}{\partial g}, \qquad N_{\sigma\sigma} = g''\frac{\partial N}{\partial g} + g'^2\frac{\partial^2 N}{\partial g^2}~,
\end{equation}
so that we can express non-Gassianity as

\begin{equation}
 \label{eq:non-Gaussianity}
 \frac{6}{5}\fnl^{\text{MP}} \simeq \frac{1}{2}\frac{\xi^2\lambda^4}{(1+\xi\lambda^2)^2}\left[\left(\frac{g''\sigma_*}{g'} - \frac{g'\sigma_*}{g}\right)\frac{1}{\lambda}
 - \frac{1}{2}\left(\frac{g'\sigma_*}{g}\right)\frac{1}{\lambda^2}\frac{\partial^2\ln \mu}{\partial (\ln g)^2}\right].
\end{equation}

\begin{figure}[h!]
\begin{center}
\includegraphics[width=0.50\textwidth]{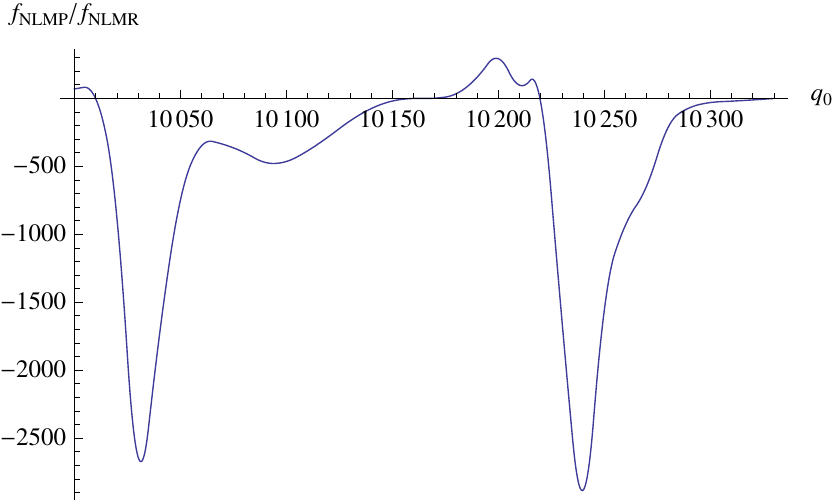}
\caption{Comparison of non-Gaussianity from modulated preheating and non-Gaussianity from modulated reheating for $w_{f}=0.25, H_0 = 0.25m$ in the limit $\sigma/M\ll1$. Here the same estimation for $\ettatext$ is used as in the previous figure.}
\label{fig:fnlMPMR}
\end{center}
\end{figure}

Note that $\fnl$ may be positive or negative depending on the parameter values. This is in contrast to the case where backreaction is negligible, studied in~\cite{Kohri:2009ac}, where $\fnl$ is always negative. The difference is due to the fact that when preheating is shut down by backreaction the end of resonance depends non-trivially on the coupling $g$. If $\sigma/M \ll 1$ the first term in the square brackets should dominate. In fact this term is of the same form as that obtained from the modulated reheating scenario. Therefore we can relate the non-Gaussianity produced by these two processes in this limit:

\begin{equation}
\fnl^{\text{MP}} \simeq -\left(\frac{2w_f}{1+w_f}\right)^3\left(\etta\right)^3 \fnl^{\text{MR}}.
\end{equation}
This relation of plotted in Figure~\ref{fig:fnlMPMR} for different values of the resonance parameter with the same estimation for $\ettatext$ as before.

\begin{figure}[!ht]
\begin{center}
\includegraphics[width=0.50\textwidth]{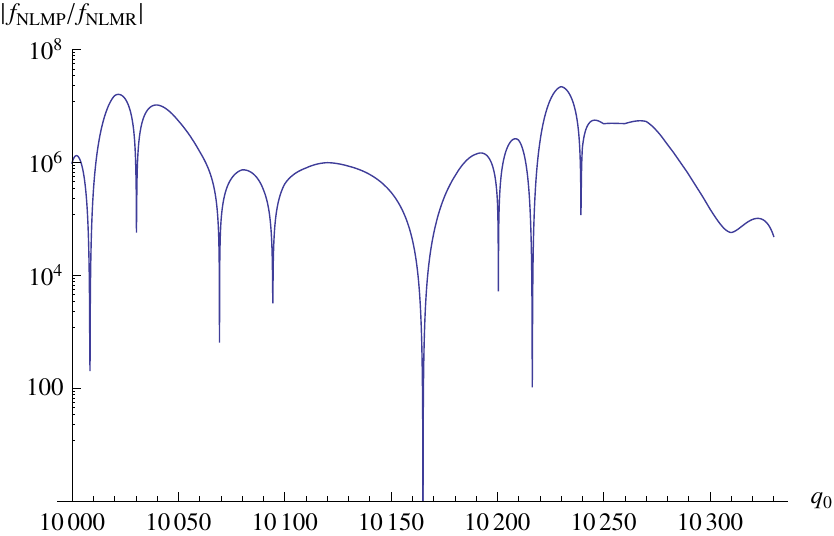}
\caption{Comparison of non-Gaussianity from modulated preheating and non-Gaussianity from modulated reheating for $w_{f}=0.25, H_0 = 0.25m$ in the limit $\sigma/M\gg1$. $\ettatext$ and $\ettatwotext$ were again estimated by solving them from the equations of motion after $15$ inflaton oscillations.}
\label{fig:fnlMPMRlargelimit}
\end{center}
\end{figure}

In the oppsite limit, $\sigma/M \gg 1$, we can can also relate the two non-Gaussianities if $\xi$ is small by solving $\xi$ in terms of $\fnl^{\text{MR}}$. We find

\begin{equation}
 \fnl^{\text{MP}} \simeq -\frac{48}{5}\left[\left(\frac{w_f}{1+w_f}\right)^3 \left(\etta\right)^3
 + 2\left(\frac{w_f}{1+w_f}\right)^2 \left(\etta\right)^2\ettatwo \right]\fnl^{\text{MR}}, \nonumber
\end{equation}
and the result is plotted in Figure~\ref{fig:fnlMPMRlargelimit}.
If $\xi$ is large, on the other hand, the $\fnl^{\text{MR}} \approx 5/2$ in the limit $\sigma/M \gg 1$ whereas non-Gaussianity from modulated preheating can still be large as long as $\xi\lambda^2 \gtrsim 1$. It is clear that in all cases modulated preheating can produce much more significant curvature perturbations and non-Gaussianities than the modulated reheating with the same coupling profile.

\section{Isocurvature perturbations}
\label{sec:isocurvature}
Let us now assume that the modulating field $\sigma$ is responsible for cold dark matter (CDM) whereas the preheat field $\chi$ will eventually give rise to radiation that consists of the Standard Model degrees of freedom.

If $\sigma$ is subdominant at the beginning of the radiation domination era when both $\phi$ and $\chi$ have decayed, we should have $\zeta\simeq \zeta_r $, where $r$ refers to radiation. The curvature perturbation in CDM is given by $\zeta_{CDM} = \zeta_{\sigma}$. On a hypersurface of uniform radiation density the isocurvature perturbation between the CDM field
$\sigma$ and radiation is from eq.~\eqref{zetai}

\begin{equation}
 \mathcal{S}_{\sigma} \equiv 3(\zeta_{\sigma} - \zeta_r) = \frac{1}{1+w_{\sigma}}\frac{\delta\rho_{\sigma}}{\rho_{\sigma}}. \label{eq:isocurvature}
\end{equation}
Assuming the potential for  $\sigma$ is harmonic with
$V(\sigma)=\frac{1}{2}m_{\sigma}^2\sigma^2$ and that the
mass $m_{\sigma}\ll H$ throughout inflation and reheating, $\sigma$ will start evolving only once the Hubble parameter has decreased to a value comparable to the mass. If $\sigma$ is the origin of CDM, we can write the CDM isocurvature perturbation as 
\begin{equation}
\mathcal{S}_{CDM} \simeq 2{\delta\sigma_*}/{\sigma_*}~, 
\end{equation}
where $\delta\sigma_*$ is the field perturbation on the initial flat hypersurface, as usual.

Using the parametrization presented in the previous section in \eqref{eq:param_definition} we can write the power spectra as

\begin{equation}
 \mathcal{P}_{\zeta} = (1+\xi\lambda^2)\mathcal{P}_{\text{inf}}, \quad \mathcal{P_{S}} = \xi \mathcal{P}_{\text{inf}}, \quad \mathcal{C_{\zeta S}} =-\xi\lambda \mathcal{P}_{\text{inf}}
\end{equation}
where $\mathcal P_{\zeta}$ and $\mathcal{P_S}$ are the curvature and isocurvature spectra respectively, and $\mathcal{C_{\zeta S}}$ the correlation spectrum. In the above we assume that the perturbations in different fields are uncorrelated at the horizon exit. The isocurvature fraction $\alpha$ and the correlation angle $\Delta$ are given by the expressions

\begin{equation}
 \label{eq:alpha}
 \alpha \equiv  \frac{\mathcal{P_S}}{\mathcal{P}_{\zeta} + \mathcal{P_S}} = \frac{\xi}{1+\xi\left(1+\lambda^2\right)}, \quad
 \cos^2\Delta \equiv  \frac{\mathcal{C}^2_{\mathcal{\zeta S}}}{\mathcal{P_{\zeta}P_{S}}} = \frac{\xi\lambda^2}{1+\xi\lambda^2}.
\end{equation}

The parameter $\alpha$ has been constrained from the CMB and large scale structure observations to be small \cite{KurkiSuonio:2004mn,Beltran:2005gr, Beltran:2005xd, Keskitalo:2006qv, Trotta:2006ww, Seljak:2006bg, Bean:2006qz, Kawasaki:2007mb, Komatsu:2008hk, Beltran:2008ei, Sollom:2009vd, Valiviita:2009bp, Komatsu:2010fb, Larson:2010gs, Valiviita:2012ub}. If the contribution from modulated preheating dominates, i.e. $\xi\lambda^2 \gg 1$, we have $\alpha = 1/(1+\lambda^2)$ and the perturbations are completely (anti-)correlated. For completely correlated perturbations\footnote{Note that our sign convention differs from that of \cite{Komatsu:2010fb,Larson:2010gs} and our correlated case corresponds to their anti-correlated case and vice versa.} $\alpha < 0.011\: (0.0047)$ from WMAP7 (WMAP7 + BAO + SN) data at 95\% confidence limit \cite{Larson:2010gs,Komatsu:2010fb}. This requires $\lambda > 10 \: (15)$ which is not achievable according to our numerical estimations for $\ettatext$. Therefore the curvature perturbation cannot be solely due to this effect. The constraint on anti-correlated case is of the same order as on the correlated case \cite{Valiviita:2012ub}.

\begin{figure}[!ht]
\begin{center}
\includegraphics[width=0.45\textwidth]{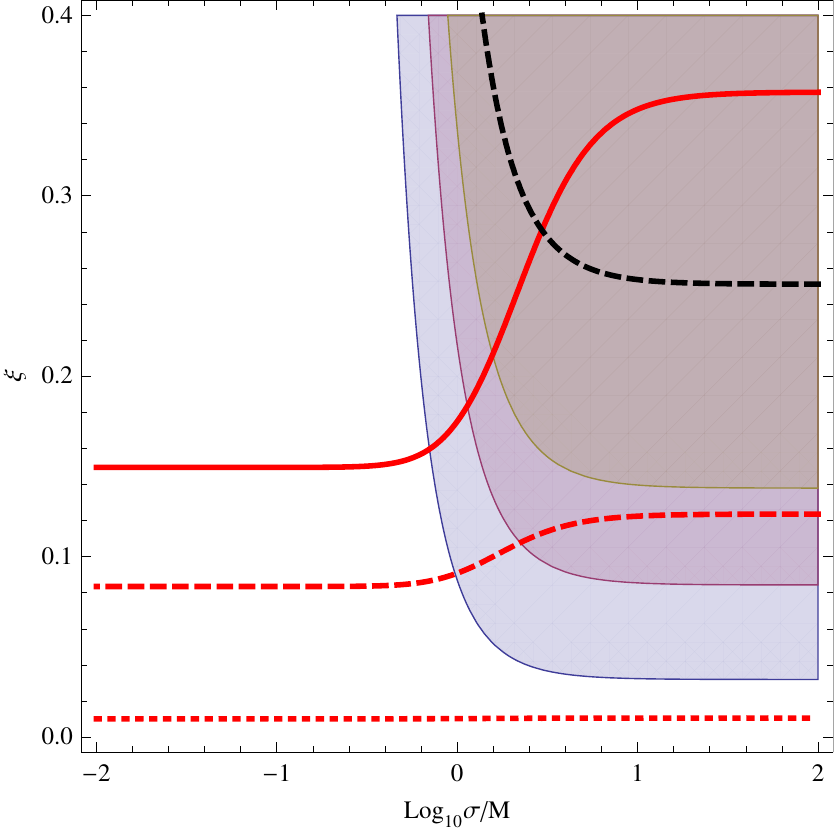}
\caption{Non-Gaussianity and isocurvature bounds for $q_0=10250$. The blue, the violet and the brown regions correspond to $\fnl>5, 25,$ and $50$ respectively. The dashed black line represents $\fnl=100$. The red curves show the isocurvature constraints with the solid, dashed and dotted curves giving respectively $\alpha = 0.13, 0.077$ and $0.01$.}
\label{fig:xi_sigma_plot}
\end{center}
\end{figure}
If modulated preheating does not significantly contribute to the curvature perturbation, $\xi\lambda \ll 1$, perturbations are uncorrelated and we have $\alpha \simeq \xi/(1+\xi)$.The constraints on isocurvature for the uncorrelated case are less stringent: $\alpha < 0.13\: (0.077)$ \cite{Komatsu:2010fb,Larson:2010gs}, and we obtain the constraint $\xi < 0.15 \: (0.084)$.

\begin{figure}[!ht]
\begin{center}
\includegraphics[width=0.45\textwidth]{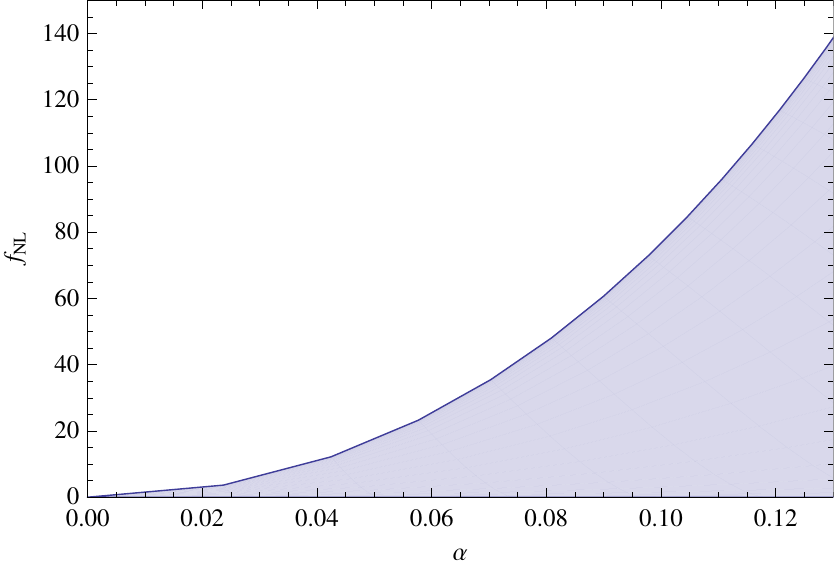}
\caption{The region of allowed non-Gaussianity and isocurvature for $q_0=10250$.}
\label{fig:fnl_alpha_plot}
\end{center}
\end{figure}

The bounds on isocurvature perturbations imply limits on the amount of non-Gaussi\-ani\-ty that can be produced. In Figure~\ref{fig:xi_sigma_plot} we show the regions of the parameter space giving significant non-Gaussianity along with the constraints from isocurvature
and in Figure~\ref{fig:fnl_alpha_plot} the dependence between non-Gaussianity and the amount of isocurvature for $q_0 =10250$, which is the value for which preheating lasts about 15 inflaton oscillations with $\Phi_0/m = 10^6$.
The general trend is the same for other values. The result is similar to that of Takahashi et al.~\cite{Takahashi:2009cx} who investigated modulated reheating with gravitino dark matter. They found that significant non-Gaussianity requires large isocurvature perturbations.

\begin{figure}[!ht]
\begin{center}
\includegraphics[width=0.45\textwidth]{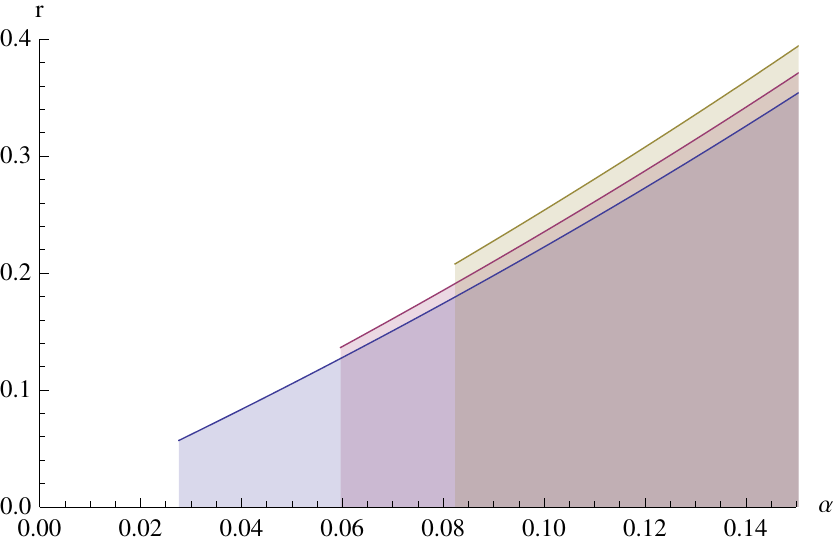}
\caption{The allowed region for $r$ and $\alpha$ for $q_0 = 10250$ with fixed $\fnl = 5,\:25,\:50$ corresponding to the blue, violet and brown shaded regions respectively. Given an observation of
$\fnl$, a bound on isocurvature implies a constraint on primordial gravitational waves. The line terminates for small $\alpha$ because small isocurvature implies small non-Gaussianity (see Fig.~\ref{fig:fnl_alpha_plot}) so that at some point the fixed $\fnl$ can no longer be achieved.}
\label{fig:r_max_alpha_plot}
\end{center}
\end{figure}

We note that if the value of the modulating field $\sigma$ at horizon crossing is required to be below the Planck mass, then from the definition~\eqref{eq:param_definition} we obtain the condition $r_* \le 2\xi$ where $r_*$ is the tensor-to-scalar ratio at horizon crossing. Thus $2\xi \equiv r_{\text{max}}$ gives an upper bound for primordial tensor perturbations. In particular, if we fix $\fnl$, we can obtain a relation between $r_{\text{max}}$ and $\alpha$. This relation is displayed in Figure~\ref{fig:r_max_alpha_plot}.
Detection of non-Gaussianty implies bounds on the amount of isocurvarture and gravitational waves. For large values of $\lambda$ the constraint on gravitational waves becomes even more stringent because $r=r_*/(1+r_*\lambda^2/2)$.

\section{Conclusions}
\label{sec:conclusions}

Motivated by the modulated reheating scenario we have investigated a model where the decay of the inflaton $\phi$ is not perturbative but happens because of resonant production of particles of the preheat field $\chi$ which is eventually shut down by backreaction of the produced particles. We have studied a simple potential of the type $g^2\phi^2\chi^2$ and have limited our analysis to preheating and have not considered the subsequent thermalisation. The spatial modulation of the inflaton coupling $g$ to the preheat field is assumed to be due to an additional scalar modulating field $\sigma$, which is light during inflation. Therefore,  during inflation the modulating field achieves entropy perturbations which are converted to curvature perturbations at preheating. We find that the contribution due to modulation to the resulting power spectrum and non-Gaussianity can be much more significant than if one assumes modulated reheating with the same set of parameters, as was discussed in section \ref{subsec:nongaussian}.

We also considered the possibility that the modulating field is CDM while the radiation in the universe is due to the dynamics of the preheat field. In this case there will be an isocurvarture perturbation in CDM which is due to the perturbation in the modulating field $\sigma$. We find that current observational constraints on isocurvature imply that the primordial curvature perturbation cannot be solely due to modulated preheating. We also find that isocurvature bounds can put severe limits on the amount of non-Gaussianity and primordial gravitational waves that can be observed. We have not performed a complete scan through parameter space but rather presented in Figure~\ref{fig:fnl_alpha_plot} an example with the resonant parameter $q_0=10250$ that demonstrates the general trend in modulated preheating, the presence of non-Gaussianity implies a constraint on the amount of isocurvature.

\acknowledgments{SR is supported by the Jenny and Antti Wihuri Foundation. KE is supported by the Academy of Finland grant 218322.}

\appendix

\section{Estimating the behavior of \texorpdfstring{\boldmath $\mu$}{mu}}
\label{sec:Floquet}

\begin{figure}[!ht]
\begin{center}
\includegraphics[width=0.50\textwidth]{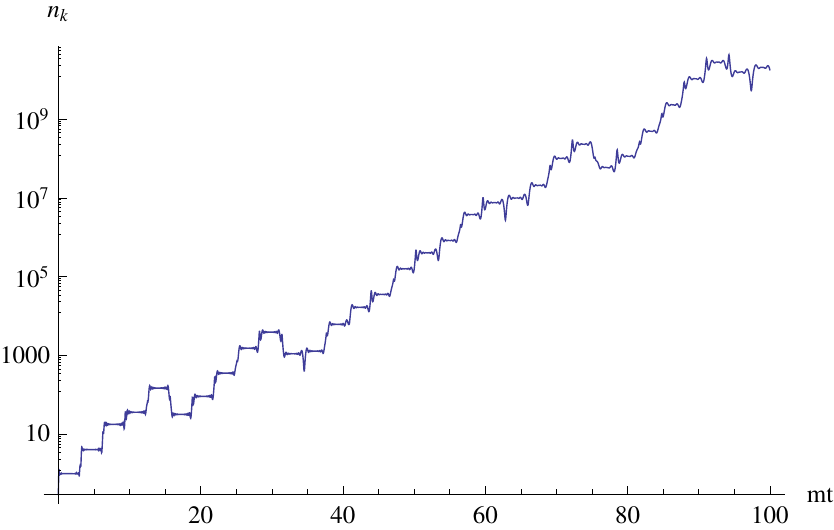}
\caption{$n_k(t)$ for $q_0 = (32\pi)^2,\:{k^2}/{m^2}\simeq 40$}
\label{fig:n_k}
\end{center}
\end{figure}

Here we estimate the magnitude of the terms $\ettatext$ and $\ettatwotext$ from the preheating dynamics. The inflaton evolves as $\phi(t) = \Phi(t)\sin(mt)$, where the amplitude decreases as $\Phi(t) = \Phi_0 a^{-3/2}(t)$. The equation of motion for the $\chi$ field (in Fourier space) is then

\begin{equation}
 \ddchi + 3H\dchi + \left[\frac{k^2}{a^2} + g^2\phi^2\right]\chik = 0 \label{eq:EOMchi}.
\end{equation}
Defining a new variable $\X(t)\equiv a^{3/2}\chik(t)$ we can rewrite the equation of motion as

\begin{equation}
 \label{eq:EOMX}
 \ddX +  \omega_k^2(t)\X = 0,
\end{equation}
where we have defined the frequency $\omega_k^2 (t) \equiv {k^2}/({m^2a^2}) + 4q \sin^2 t$ with the resonance parameter $q \equiv {g^2\Phi^2}/({4m^2})$ . Time is now measured in units of $m^{-1}$. The occupation number for particles of mode $\mathbf k$ is given by

\begin{equation}
 n_k = \frac{\omega_{k}}{2}\left[\frac{|\dX|^2}{\omega_k^2} + |\X|^2\right] - \frac{1}{2}.
\end{equation}
As the inflaton goes through the minimum of the potential, the particle density jumps rapidly either up or down, depending on the phase acquired since the last oscillation. Density is twice as likely to increase than to decrease so that  overall $n_k$ grows. The behavior of $n_k$ is plotted in Figure~\ref{fig:n_k}.

We define an effective Floquet exponent $\mu_k$ as

\begin{equation}
 \mu_k (t) \equiv \frac{1}{2mt}\ln(2n_k).
\end{equation}
The number density of $\chi$ particles is then obtained by integrating over all the modes

\begin{equation}
 \label{eq:n_chi_integral}
 n_{\chi}(t) = \frac{1}{(2\pi a)^3}\int\ud^3 k n_k(t) = \frac{1}{4\pi^2a^3}\int\ud k k^2 e^{2\mu_k mt}.
\end{equation}
For late times the integral can be evaluated using the saddle-point (Lagrange) method, which gives

\begin{equation}
 \label{eq:n_chi_mu''}
 n_{\chi}(t) \simeq \frac{1}{4\pi^2 a^3}\left[\sqrt{\frac{\pi}{m t |\mu_k''|}}k^2 e^{2\mu_k m t}\right]_{k=k_m},
\end{equation}
where $'$ denotes a derivative with respect to $k$ and $k_m$ is the value of $k$ that maximizes $\mu_k$, and $\mu_{k=k_m}$ is the effective Floquet exponent $\mu$ used in Section~\ref{sec:modpre}. We can estimate $\mu \simeq \frac{1}{2}|\mu''|(\Delta k)^2$ where $\Delta k$ is the width of the resonance band, which can be estimated as $\Delta k \sim \sqrt{2/\pi}q_0^{1/4} m$ (see \cite{Kofman:1997yn}). Taking the $k_m$ to be in the middle of the resonance band, $k_m \sim  \frac{1}{2}\sqrt{2/\pi}q_0^{1/4} m$, we obtain equation~\eqref{eq:n_chi}.

\begin{figure}[!ht]
\begin{center}
\includegraphics[width=0.50\textwidth]{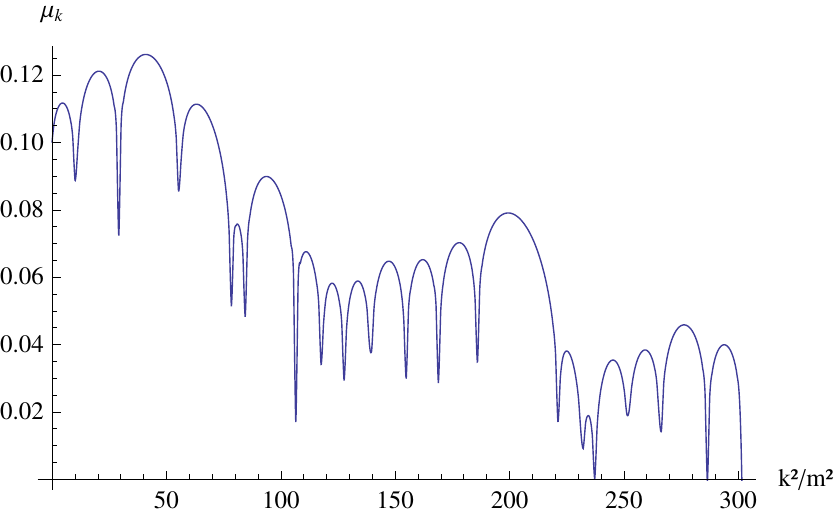}
\caption{\label{fig:mu_k}$\mu_k$ as a function of $k^2/m^2$ for $q_0 = (32\pi)^2$ after $15$ inflaton oscillations.}
\end{center}
\end{figure}

\begin{figure}[!ht]
\begin{center}
\includegraphics[width=0.50\textwidth]{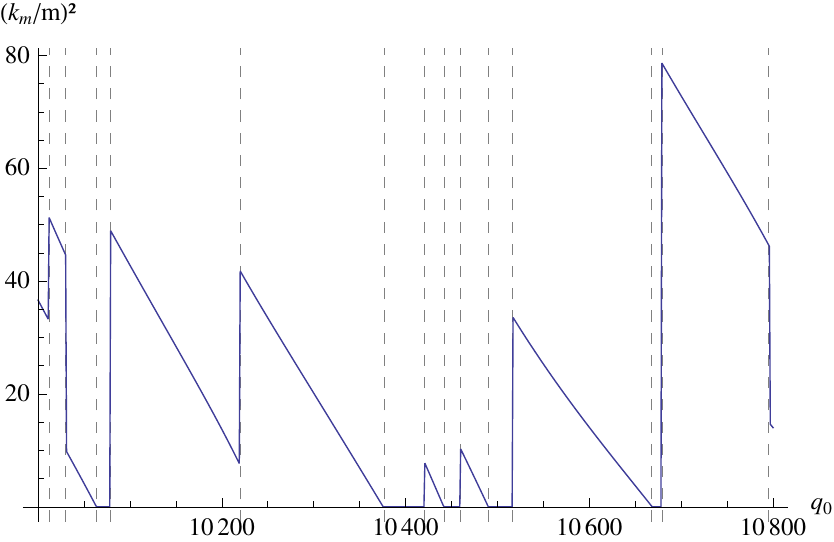}
\caption{\label{fig:k2max_q0_10000}The location of the global maximum of $\mu_k$ as a function of $q_0$. The dashed vertical lines correspond to places where different peaks take over and the places where the maximum is at the origin. In the vicinity of these points the analytic estimates are not reliable.}
\end{center}
\end{figure}

\begin{figure}[!ht]
\begin{center}
\includegraphics[width=0.50\textwidth]{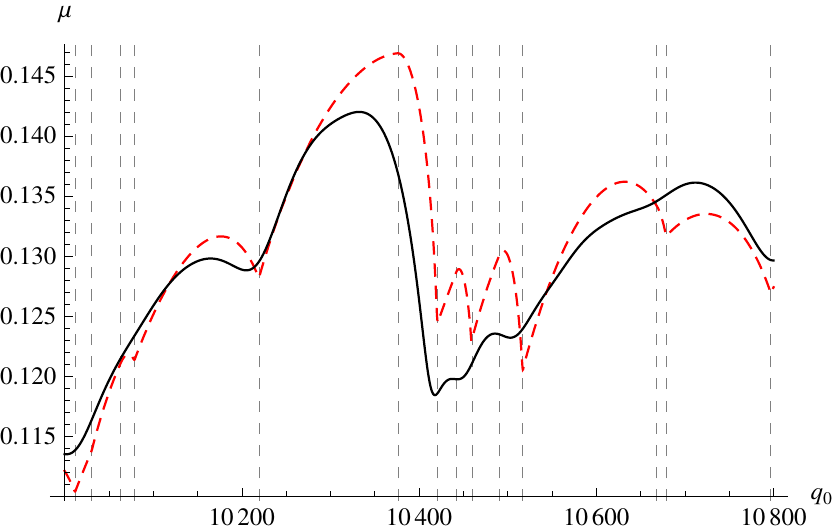}
\caption{\label{fig:mu_q0_10000}$\mu$ as a function of $q_0$ after $15$ inflaton oscillations. The dashed red curve corresponds to calculating the global maximum of $\mu_k$ and the solid black curve is obtained by using eq.~\eqref{eq:n_chi} as the the definition of $\mu$ and solving for it from the integrated $n_{\chi}$.}
\end{center}
\end{figure}

Figure~\ref{fig:mu_k} shows the behavior of $\mu_k (t)$ as a function of $k^2/m^2$ for $q_0 = (32\pi)^2$ after $15$ inflaton oscillations.
As $q_0$ is changed the positions and magnitudes of the peaks change continuously and therefore so does the global maximum. However the change is not monotonic because as the peak giving the global maximum goes down another peak may grow past it and become the new global maximum. Figures~\ref{fig:k2max_q0_10000} and~\ref{fig:mu_q0_10000} show how the global maximum changes with $q_0$.

\begin{figure}[h!]
\subfloat[][]{
\includegraphics[width=0.48\textwidth]{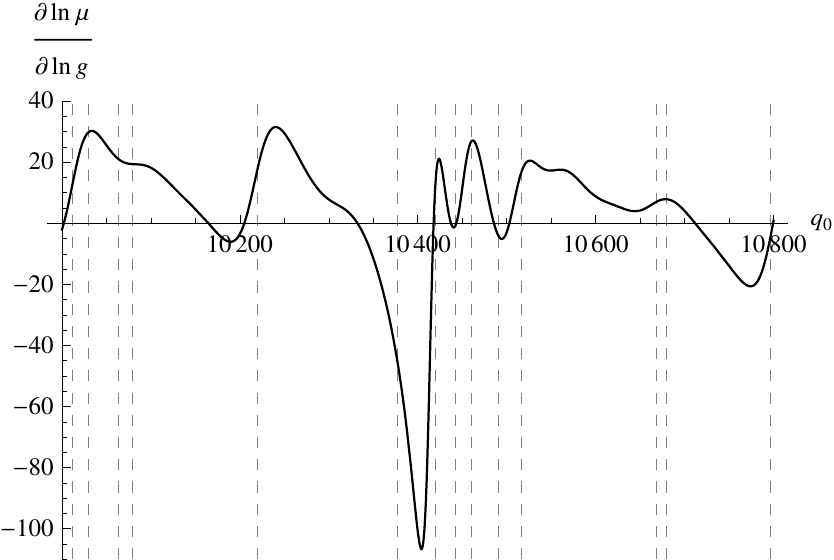}}
\quad
\subfloat[][]{
\includegraphics[width=0.48\textwidth]{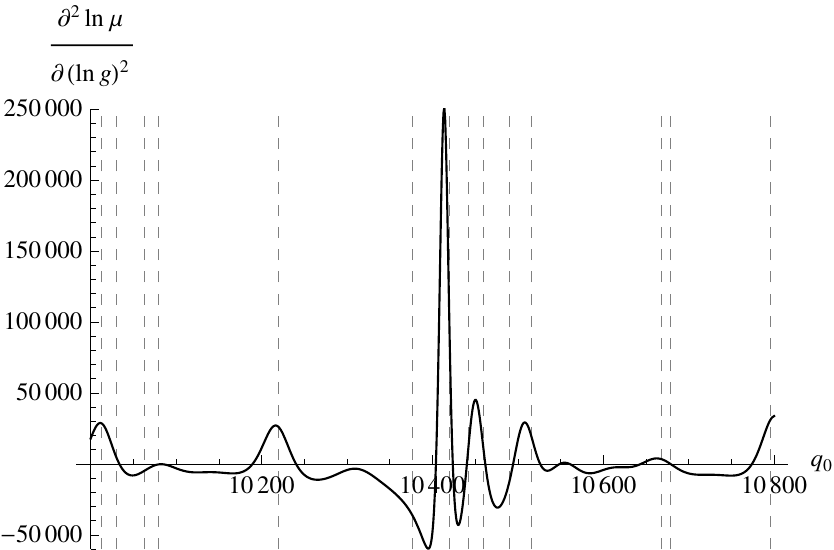}}
\caption{$\ettatext$ (a) and $\ettatwotext$ (b) calculated after $15$ oscillation of the inflaton for $q\sim10^4$}
\label{fig:eta12_q0_10000}
\end{figure}

The analytic estimates given above are not reliable in the vicinity of the transition from one peak to the next because at that point there are two separate peaks giving the same maximum and the saddle point approximation is not applicable. The estimates are also unreliable for the cases where the global maximum is at $k=0$ as can be seen from eq.~\eqref{eq:n_chi_mu''}. For this reason, we numerically obtain the number density $n_{\chi}$ by calculating the integral~\eqref{eq:n_chi_integral} and use eq.~\eqref{eq:n_chi} as the definition of $\mu$. This way we obtain the correct behavior for all values. The behavior of $\ettatext$ and $\ettatwotext$ after 15 oscillations of the inflaton is shown in Figures~\ref{fig:eta12_q0_10000}.

%
%

\bibliographystyle{JHEP.bst}
\bibliography{noteref}

\providecommand{\href}[2]{#2}\begingroup\raggedright\begin{thebibliography}{10}

\bibitem{Abbott:1982hn}
L.~Abbott, E.~Farhi, and M.~B. Wise, {\it {Particle Production in the New
  Inflationary Cosmology}},  {\em Phys.Lett.} {\bf B117} (1982) 29.

\bibitem{Dolgov:1982th}
A.~Dolgov and A.~D. Linde, {\it {Baryon Asymmetry in Inflationary Universe}},
  {\em Phys.Lett.} {\bf B116} (1982) 329.

\bibitem{Albrecht:1982mp}
A.~Albrecht, P.~J. Steinhardt, M.~S. Turner, and F.~Wilczek, {\it {Reheating an
  Inflationary Universe}},  {\em Phys.Rev.Lett.} {\bf 48} (1982) 1437.

\bibitem{Dvali:2003em}
G.~Dvali, A.~Gruzinov, and M.~Zaldarriaga, {\it {A new mechanism for generating
  density perturbations from inflation}},  {\em Phys.Rev.} {\bf D69} (2004)
  023505, [\href{http://xxx.lanl.gov/abs/astro-ph/0303591}{{\tt
  astro-ph/0303591}}].

\bibitem{Dvali:2003ar}
G.~Dvali, A.~Gruzinov, and M.~Zaldarriaga, {\it {Cosmological perturbations
  from inhomogeneous reheating, freezeout, and mass domination}},  {\em
  Phys.Rev.} {\bf D69} (2004) 083505,
  [\href{http://xxx.lanl.gov/abs/astro-ph/0305548}{{\tt astro-ph/0305548}}].

\bibitem{Zaldarriaga:2003my}
M.~Zaldarriaga, {\it {Non-Gaussianities in models with a varying inflaton decay
  rate}},  {\em Phys.Rev.} {\bf D69} (2004) 043508,
  [\href{http://xxx.lanl.gov/abs/astro-ph/0306006}{{\tt astro-ph/0306006}}].

\bibitem{Traschen:1990sw}
J.~H. Traschen and R.~H. Brandenberger, {\it {PARTICLE PRODUCTION DURING
  OUT-OF-EQUILIBRIUM PHASE TRANSITIONS}},  {\em Phys.Rev.} {\bf D42} (1990)
  2491--2504.

\bibitem{Kofman:1994rk}
L.~Kofman, A.~D. Linde, and A.~A. Starobinsky, {\it {Reheating after
  inflation}},  {\em Phys. Rev. Lett.} {\bf 73} (1994) 3195--3198,
  [\href{http://xxx.lanl.gov/abs/hep-th/9405187}{{\tt hep-th/9405187}}].

\bibitem{Kofman:1997yn}
L.~Kofman, A.~D. Linde, and A.~A. Starobinsky, {\it {Towards the theory of
  reheating after inflation}},  {\em Phys.Rev.} {\bf D56} (1997) 3258--3295,
  [\href{http://xxx.lanl.gov/abs/hep-ph/9704452}{{\tt hep-ph/9704452}}].

\bibitem{Bassett:2005xm}
B.~A. Bassett, S.~Tsujikawa, and D.~Wands, {\it {Inflation dynamics and
  reheating}},  {\em Rev.Mod.Phys.} {\bf 78} (2006) 537--589,
  [\href{http://xxx.lanl.gov/abs/astro-ph/0507632}{{\tt astro-ph/0507632}}].

\bibitem{Allahverdi:2010xz}
R.~Allahverdi, R.~Brandenberger, F.-Y. Cyr-Racine, and A.~Mazumdar, {\it
  {Reheating in Inflationary Cosmology: Theory and Applications}},  {\em
  Ann.Rev.Nucl.Part.Sci.} {\bf 60} (2010) 27--51,
  [\href{http://xxx.lanl.gov/abs/1001.2600}{{\tt arXiv:1001.2600}}].

\bibitem{Kofman:2003nx}
L.~Kofman, {\it {Probing string theory with modulated cosmological
  fluctuations}},  \href{http://xxx.lanl.gov/abs/astro-ph/0303614}{{\tt
  astro-ph/0303614}}.

\bibitem{Podolsky:2005bw}
D.~I. Podolsky, G.~N. Felder, L.~Kofman, and M.~Peloso, {\it {Equation of state
  and beginning of thermalization after preheating}},  {\em Phys. Rev.} {\bf
  D73} (2006) 023501, [\href{http://xxx.lanl.gov/abs/hep-ph/0507096}{{\tt
  hep-ph/0507096}}].

\bibitem{Kohri:2009ac}
K.~Kohri, D.~H. Lyth, and C.~A. Valenzuela-Toledo, {\it {Preheating and the
  non-gaussianity of the curvature perturbation}},  {\em JCAP} {\bf 1002}
  (2010) 023 [{\em Erratum ibid.} {\bf 1009} (2011) E01] [\href{http://xxx.lanl.gov/abs/0904.0793}{{\tt
  arXiv:0904.0793}}].

\bibitem{Bernardeau:2004zz}
F.~Bernardeau, L.~Kofman, and J.-P. Uzan, {\it {Modulated fluctuations from
  hybrid inflation}},  {\em Phys.Rev.} {\bf D70} (2004) 083004,
  [\href{http://xxx.lanl.gov/abs/astro-ph/0403315}{{\tt astro-ph/0403315}}].

\bibitem{Matsuda:2007tr}
T.~Matsuda, {\it {Cosmological perturbations from inhomogeneous preheating and
  multi-field trapping}},  {\em JHEP} {\bf 0707} (2007) 035,
  [\href{http://xxx.lanl.gov/abs/0707.0543}{{\tt arXiv:0707.0543}}].

\bibitem{GarciaBellido:1995qq}
J.~Garcia-Bellido and D.~Wands, {\it {Metric perturbations in two field
  inflation}},  {\em Phys.Rev.} {\bf D53} (1996) 5437--5445,
  [\href{http://xxx.lanl.gov/abs/astro-ph/9511029}{{\tt astro-ph/9511029}}].

\bibitem{Wands:2000dp}
D.~Wands, K.~A. Malik, D.~H. Lyth, and A.~R. Liddle, {\it {A New approach to
  the evolution of cosmological perturbations on large scales}},  {\em
  Phys.Rev.} {\bf D62} (2000) 043527,
  [\href{http://xxx.lanl.gov/abs/astro-ph/0003278}{{\tt astro-ph/0003278}}].

\bibitem{Gordon:2000hv}
C.~Gordon, D.~Wands, B.~A. Bassett, and R.~Maartens, {\it {Adiabatic and
  entropy perturbations from inflation}},  {\em Phys.Rev.} {\bf D63} (2001)
  023506, [\href{http://xxx.lanl.gov/abs/astro-ph/0009131}{{\tt
  astro-ph/0009131}}].

\bibitem{Enqvist:2001zp}
K.~Enqvist and M.~S. Sloth, {\it {Adiabatic CMB perturbations in pre - big bang
  string cosmology}},  {\em Nucl.Phys.} {\bf B626} (2002) 395--409,
  [\href{http://xxx.lanl.gov/abs/hep-ph/0109214}{{\tt hep-ph/0109214}}].

\bibitem{Lyth:2001nq}
D.~H. Lyth and D.~Wands, {\it {Generating the curvature perturbation without an
  inflaton}},  {\em Phys.Lett.} {\bf B524} (2002) 5--14,
  [\href{http://xxx.lanl.gov/abs/hep-ph/0110002}{{\tt hep-ph/0110002}}].

\bibitem{Moroi:2001ct}
T.~Moroi and T.~Takahashi, {\it {Effects of cosmological moduli fields on
  cosmic microwave background}},  {\em Phys.Lett.} {\bf B522} (2001) 215--221 [{\em Erratum ibid.} {\bf B 539} (2002) 303] 
  [\href{http://xxx.lanl.gov/abs/hep-ph/0110096}{{\tt hep-ph/0110096}}].

\bibitem{Mazumdar:2010sa}
A.~Mazumdar and J.~Rocher, {\it {Particle physics models of inflation and
  curvaton scenarios}},  {\em Phys.Rept.} {\bf 497} (2011) 85--215,
  [\href{http://xxx.lanl.gov/abs/1001.0993}{{\tt arXiv:1001.0993}}].

\bibitem{Bardeen:1983qw}
J.~M. Bardeen, P.~J. Steinhardt, and M.~S. Turner, {\it {Spontaneous Creation
  of Almost Scale - Free Density Perturbations in an Inflationary Universe}},
  {\em Phys.Rev.} {\bf D28} (1983) 679.

\bibitem{Bardeen:1989}
J.~M. Bardeen, {\em \emph{in} Particle Physics and Cosmology}.
\newblock Gordon and Breach, New York, 1989.

\bibitem{Lyth:2004gb}
D.~H. Lyth, K.~A. Malik, and M.~Sasaki, {\it {A General proof of the
  conservation of the curvature perturbation}},  {\em JCAP} {\bf 0505} (2005)
  004, [\href{http://xxx.lanl.gov/abs/astro-ph/0411220}{{\tt
  astro-ph/0411220}}].

\bibitem{Frolov:2008hy}
A.~V. Frolov, {\it {DEFROST: A New Code for Simulating Preheating after
  Inflation}},  {\em JCAP} {\bf 0811} (2008) 009,
  [\href{http://xxx.lanl.gov/abs/0809.4904}{{\tt arXiv:0809.4904}}].

\bibitem{Sainio:2009hm}
J.~Sainio, {\it {CUDAEASY - a GPU Accelerated Cosmological Lattice Program}},
  {\em Comput.Phys.Commun.} {\bf 181} (2010) 906--912,
  [\href{http://xxx.lanl.gov/abs/0911.5692}{{\tt arXiv:0911.5692}}].

\bibitem{Easther:2010qz}
R.~Easther, H.~Finkel, and N.~Roth, {\it {PSpectRe: A Pseudo-Spectral Code for
  (P)reheating}},  {\em JCAP} {\bf 1010} (2010) 025,
  [\href{http://xxx.lanl.gov/abs/1005.1921}{{\tt arXiv:1005.1921}}].

\bibitem{Micha:2002ey}
R.~Micha and I.~I. Tkachev, {\it {Relativistic turbulence: A Long way from
  preheating to equilibrium}},  {\em Phys.Rev.Lett.} {\bf 90} (2003) 121301,
  [\href{http://xxx.lanl.gov/abs/hep-ph/0210202}{{\tt hep-ph/0210202}}].

\bibitem{Micha:2004bv}
R.~Micha and I.~I. Tkachev, {\it {Turbulent thermalization}},  {\em Phys.Rev.}
  {\bf D70} (2004) 043538, [\href{http://xxx.lanl.gov/abs/hep-ph/0403101}{{\tt
  hep-ph/0403101}}].

\bibitem{KurkiSuonio:2004mn}
H.~Kurki-Suonio, V.~Muhonen, and J.~Valiviita, {\it {Correlated Primordial
  Perturbations in Light of CMB and LSS Data}},  {\em Phys. Rev.} {\bf D71}
  (2005) 063005, [\href{http://xxx.lanl.gov/abs/astro-ph/0412439}{{\tt
  astro-ph/0412439}}].

\bibitem{Beltran:2005gr}
M.~Beltran, J.~Garcia-Bellido, J.~Lesgourgues, and M.~Viel, {\it {Squeezing the
  window on isocurvature modes with the Lyman- alpha forest}},  {\em Phys.
  Rev.} {\bf D72} (2005) 103515,
  [\href{http://xxx.lanl.gov/abs/astro-ph/0509209}{{\tt astro-ph/0509209}}].

\bibitem{Beltran:2005xd}
M.~Beltran, J.~Garcia-Bellido, J.~Lesgourgues, A.~R. Liddle, and A.~Slosar,
  {\it {Bayesian model selection and isocurvature perturbations}},  {\em Phys.
  Rev.} {\bf D71} (2005) 063532,
  [\href{http://xxx.lanl.gov/abs/astro-ph/0501477}{{\tt astro-ph/0501477}}].

\bibitem{Keskitalo:2006qv}
R.~Keskitalo, H.~Kurki-Suonio, V.~Muhonen, and J.~Valiviita, {\it {Hints of
  Isocurvature Perturbations in the Cosmic Microwave Background?}},  {\em JCAP}
  {\bf 0709} (2007) 008, [\href{http://xxx.lanl.gov/abs/astro-ph/0611917}{{\tt
  astro-ph/0611917}}].

\bibitem{Trotta:2006ww}
R.~Trotta, {\it {The isocurvature fraction after WMAP 3-year data}},  {\em Mon.
  Not. Roy. Astron. Soc. Lett.} {\bf 375} (2007) L26--L30,
  [\href{http://xxx.lanl.gov/abs/astro-ph/0608116}{{\tt astro-ph/0608116}}].

\bibitem{Seljak:2006bg}
U.~Seljak, A.~Slosar, and P.~McDonald, {\it {Cosmological parameters from
  combining the Lyman-alpha forest with CMB, galaxy clustering and SN
  constraints}},  {\em JCAP} {\bf 0610} (2006) 014,
  [\href{http://xxx.lanl.gov/abs/astro-ph/0604335}{{\tt astro-ph/0604335}}].

\bibitem{Bean:2006qz}
R.~Bean, J.~Dunkley, and E.~Pierpaoli, {\it {Constraining Isocurvature Initial
  Conditions with WMAP 3- year data}},  {\em Phys. Rev.} {\bf D74} (2006)
  063503, [\href{http://xxx.lanl.gov/abs/astro-ph/0606685}{{\tt
  astro-ph/0606685}}].

\bibitem{Kawasaki:2007mb}
M.~Kawasaki and T.~Sekiguchi, {\it {Cosmological Constraints on Isocurvature
  and Tensor Perturbations}},  {\em Prog. Theor. Phys.} {\bf 120} (2008)
  995--1016, [\href{http://xxx.lanl.gov/abs/0705.2853}{{\tt arXiv:0705.2853}}].

\bibitem{Komatsu:2008hk}
{\bf WMAP} Collaboration, E.~Komatsu et~al., {\it {Five-Year Wilkinson
  Microwave Anisotropy Probe (WMAP) Observations: Cosmological
  Interpretation}},  {\em Astrophys.J.Suppl.} {\bf 180} (2009) 330--376,
  [\href{http://xxx.lanl.gov/abs/0803.0547}{{\tt arXiv:0803.0547}}].

\bibitem{Beltran:2008ei}
M.~Beltran, {\it {Isocurvature, non-gaussianity and the curvaton model}},  {\em
  Phys. Rev.} {\bf D78} (2008) 023530,
  [\href{http://xxx.lanl.gov/abs/0804.1097}{{\tt arXiv:0804.1097}}].

\bibitem{Sollom:2009vd}
I.~Sollom, A.~Challinor, and M.~P. Hobson, {\it {Cold Dark Matter Isocurvature
  Perturbations: Constraints and Model Selection}},  {\em Phys. Rev.} {\bf D79}
  (2009) 123521, [\href{http://xxx.lanl.gov/abs/0903.5257}{{\tt
  arXiv:0903.5257}}].

\bibitem{Valiviita:2009bp}
J.~Valiviita and T.~Giannantonio, {\it {Constraints on primordial isocurvature
  perturbations and spatial curvature by Bayesian model selection}},  {\em
  Phys.Rev.} {\bf D80} (2009) 123516,
  [\href{http://xxx.lanl.gov/abs/0909.5190}{{\tt arXiv:0909.5190}}].

\bibitem{Komatsu:2010fb}
{\bf WMAP} Collaboration, E.~Komatsu et~al., {\it {Seven-Year Wilkinson
  Microwave Anisotropy Probe (WMAP) Observations: Cosmological
  Interpretation}},  {\em Astrophys.J.Suppl.} {\bf 192} (2011) 18,
  [\href{http://xxx.lanl.gov/abs/1001.4538}{{\tt arXiv:1001.4538}}].

\bibitem{Larson:2010gs}
D.~Larson, J.~Dunkley, G.~Hinshaw, E.~Komatsu, M.~Nolta, et~al., {\it
  {Seven-Year Wilkinson Microwave Anisotropy Probe (WMAP) Observations: Power
  Spectra and WMAP-Derived Parameters}},  {\em Astrophys.J.Suppl.} {\bf 192}
  (2011) 16, [\href{http://xxx.lanl.gov/abs/1001.4635}{{\tt arXiv:1001.4635}}].

\bibitem{Valiviita:2012ub}
J.~Valiviita, M.~Savelainen, M.~Talvitie, H.~Kurki-Suonio, and S.~Rusak, {\it
  {Constraints on scalar and tensor perturbations in phenomenological and
  two-field inflation models: Bayesian evidences for primordial isocurvature
  and tensor modes}},  {\em ApJ} {\bf 753} (2012) 151,
  [\href{http://xxx.lanl.gov/abs/1202.2852}{{\tt arXiv:1202.2852}}].

\bibitem{Takahashi:2009cx}
T.~Takahashi, M.~Yamaguchi, and S.~Yokoyama, {\it {Primordial Non-Gaussianity
  in Models with Dark Matter Isocurvature Fluctuations}},  {\em Phys.Rev.} {\bf
  D80} (2009) 063524, [\href{http://xxx.lanl.gov/abs/0907.3052}{{\tt
  arXiv:0907.3052}}].

\end{thebibliography}\endgroup

\end{document}